\documentclass[aps,twocolumn,pra,superscriptaddress,showpacs]{revtex4-1}
\usepackage{newlfont}
\usepackage{color}           
\usepackage{soul}            
\usepackage{amssymb}
\usepackage{amsfonts}
\usepackage{amsmath}
\usepackage{wasysym}
\usepackage{graphicx}
\usepackage{amsthm}
\usepackage{bm}
\usepackage[colorlinks=true, citecolor=blue, urlcolor=blue ]{hyperref}
\usepackage{times,txfonts}

\begin{document}


\title{Characterizing non-Markovianity via quantum interferometric power}

\author{Himadri Shekhar Dhar}
\affiliation{
Harish-Chandra Research Institute, Chhatnag Road, Jhunsi, Allahabad 211 019, India}
\affiliation{School of Physical Sciences, Jawaharlal Nehru University, New Delhi 110 067, India}
\author{Manabendra Nath Bera}
\affiliation{
Harish-Chandra Research Institute, Chhatnag Road, Jhunsi, Allahabad 211 019, India}
\author{Gerardo Adesso}
\affiliation{School of Mathematical Sciences, \\The University of Nottingham, University Park,
Nottingham NG7 2RD, United Kingdom}

\begin{abstract}
Non-Markovian evolution in open quantum systems is often characterized in terms of the backflow of information from environment to system and is thus an important facet in investigating the performance and robustness of quantum information protocols. In this work, we explore non-Markovianity through the breakdown of monotonicity of a metrological figure of merit, called the quantum interferometric power,
which is based on the minimal quantum Fisher information obtained by local unitary evolution of one part of the system, and can be interpreted as a quantifier of quantum correlations beyond entanglement. We investigate our proposed non-Markovianity indicator in two relevant examples. First, we consider the action of a single-party dephasing channel on a maximally entangled two-qubit state, by applying the Jamio{\l}kowski-Choi isomorphism. We observe that the proposed measure is consistent with established non-Markovianity quantifiers  defined using other approaches based on dynamical divisibility, distinguishability, and breakdown of monotonicity for the quantum mutual information.
Further, we consider the dynamics of two-qubit Werner states, under the action of a local, single-party amplitude damping channel, and observe that the nonmonotonic evolution of the quantum interferometric power is more robust than the corresponding one for entanglement in
capturing the backflow of quantum information associated with the non-Markovian process. Implications for the role of non-Markovianity in quantum metrology and possible extensions to continuous variable systems are discussed.
%

\end{abstract}

\maketitle

\section{Introduction}
\label{intro}

\noindent The fundamental essence of any information protocol resides in the amount of accessible information in a given physical system.
In a realistic quantum process, the system is rarely isolated and commonly interacts with an environment, which results
in the information being scattered in the typically large Hilbert space of the environment.
In most theoretical models, an open system is studied through the dynamics of the reduced density matrix, upon tracing over the environmental degrees of freedom \cite{open1,open2}. The dynamics of the system is described by completely positive semigroup maps, or equivalently by the solution of a master equation in Lindblad form \cite{lin1,lin2}. This formalism assumes weak system-environment coupling, short environment correlation time and a ``memoryless" transfer of information from the system to the environment leading to information erasure \cite{open1,gorini}. Such a dynamical model for open quantum systems is called Markovian. However,
it is readily observed that the Markovian formalism is not always optimal or justified when dealing with many important open quantum systems, especially in complex biological models or interacting many-body systems in condensed matter physics \cite{RHP}.
More accurately, the system-environment interaction needs to be treated as non-Markovian, which tends to deviate from the completely positive semigroup dynamics \cite{wolf,RHP}, thus making the corresponding mathematical formalism difficult. Incidentally, non-Markovian dynamics are not ``memoryless" and allow a backflow of information from the environment \cite{BLP,mach} (cf.~\cite{franco}) to the system, a fact which has interesting ramifications from the perspective of quantum information theory \cite{qit}. For example, non-Markovian processes have been shown to preserve entanglement \cite{ent} in many-body \cite{qmb} and biomolecular \cite{qbio} systems, and have been exploited in quantum key distribution \cite{qkd}, enhancing precision in quantum metrology \cite{qmet}, and implementing certain quantum information protocols \cite{qprot,capacity}. Non-Markovianity also plays a detrimental role in quantum Darwinism, thus impeding the emergence of classical objectivity from a quantum world \cite{darwin}.

With recent development of experimental techniques to engineer and control system-environment interactions \cite{control} (see also Ref.~\cite{open1}), there is considerable interest in characterizing and quantifying non-Markovian dynamics and investigating possible applications in scalable quantum technologies \cite{qkd,qmet,qprot,capacity} that are robust against environment-induced decoherence \cite{qmb} or phenomena such as entanglement sudden death \cite{esd}.
Although the concept of non-Markovianity is well established in the classical realm \cite{mark}, its quantum extension is often riddled with inconsistency and subtle variations. This has led to a substantial amount of literature attempting to quantitatively characterize non-Markovianity based primarily on the nonmonotonic time evolution of some quantum information measure (for reviews, see \cite{qmark,rep,arxiv}). Such nonmonotonic behavior arises from the nondivisibility of the completely positive and trace preserving (CPTP) maps \cite{RHP} that describe the dynamics of the open quantum system, which is perhaps the most established marker of non-Markovianity \cite{rep,arxiv,jpb} (cf. \cite{anti}).
The nondivisibility of a CPTP map is necessary for the occurrence of information backflow from the environment or the presence of the environment memory \cite{BLP}.

\begin{figure}
\centering
\includegraphics[width=0.45\textwidth, angle=0]{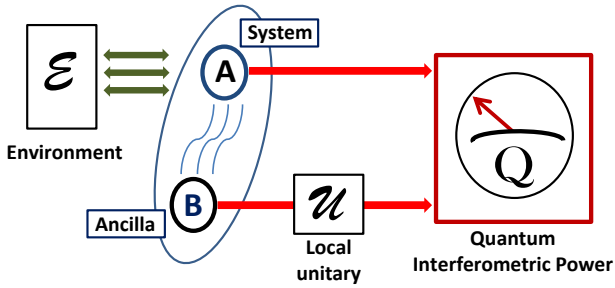}
\caption{\label{fig:0}(Color online). A scheme showing the protocol under consideration. The system $a$ interacts with the environment, while the quantum interferometric power is computed by applying local unitaries on the ancilla part $b$, which serves as the measuring apparatus.  }
\end{figure}

A number of non-Markovian measures and witnesses have been proposed. Among the most important ones, let us mention those based on the deviation of the dynamical maps from divisible CPTP maps \cite{wolf,RHP} and those based on the nonmonotonicity of the trace distance or distinguishability \cite{BLP}, entanglement \cite{RHP}, quantum mutual information \cite{LUO}, and channel capacities \cite{capacity}. Other significant attempts to quantify non-Markovianity include the flow of quantum Fisher information \cite{Fish}, fidelity between dynamical time-evolved states \cite{AKR}, distinguishability in Gaussian quantum systems in terms of fidelity \cite{gaus}, volume of Gaussian states \cite{gaus2}, backflow via accessible information \cite{fan}, and local quantum uncertainty \cite{locunc}.
Recent proposals have also been made to characterize non-Markovianity in direct analogy to entanglement theory \cite{newmanis} and to study the links between system-environment quantum correlations and non-Markovian behavior \cite{smirne}.
Interestingly, although the different non-Markovian measures and witnesses emanate from the dynamical divisibility criteria, the inverse implication is not always true, which makes them incompatible with each other for general open system dynamics \cite{qmark,rep,incompat} (cf.~\cite{LUO, compat}).

In this work, we propose to characterize the non-Markovianity of an open system evolution through the nonmonotonic behavior of a quantum metrological figure of merit, called the quantum interferometric power (QIP) \cite{ades, mb},
which is defined in terms of the minimal quantum Fisher information \cite{QFI} obtained by local unitary evolution of one part of the system. The QIP is an important information-theoretic tool that also quantifies discordlike quantum correlations in a bipartite system \cite{ades, mb} and is related to the minimum speed of evolution of a quantum system in the projective Hilbert space \cite{mb}.
To capture the non-Markovianity in open quantum evolutions, we consider a single qubit  (say, $a$) as the principal  \emph{system}, interacting with an environment. A second qubit (say, $b$) plays the role of an ancilla.
We consider the action of the environment on the system in terms of the dephasing and amplitude damping channels. Using the Jamio{\l}kowski-Choi isomorphism \cite{choi},  single-qubit operations can be used to study the bipartite (system + ancilla) behavior. The QIP of the system is measured by applying local unitaries on the ancilla, which acts as a measuring apparatus. The non-Markovianity of the evolution is characterized by quantifying the nonmonotonic behavior of the QIP.
The paper is organized as follows. In Sec. II, we briefly present the concept and definition of the QIP and introduce a non-Markovianity measure based on its nonmonotonic evolution. In Sec. III, we consider a prototypical single-qubit dephasing model for the two-qubit (system + ancilla) state, and show that the non-Markovianity measure derived using QIP is qualitatively consistent with measures based on distinguishabilty, divisibilty, and quantum mutual information. In Sec. IV, we consider a single-qubit amplitude damping model and investigate the flow of QIP in the non-Markovian regime. We observe that the measure appropriately captures the backflow of information and
is more robust compared to a non-Markovianity measure based on entanglement. We discuss the results, possible extensions, and potential benefits of the introduced non-Markovianity indicator in Sec. V.

\section{Characterizing non-Markovianity via quantum interferometric power}
\label{gmq}

\subsection{Quantum interferometric power}

The QIP is a metrological figure of merit that quantifies the guaranteed precision enabled by a bipartite probe state for the task of black-box quantum parameter estimation \cite{ades,mb,GIP,Gmb}. Let us consider a bipartite, system + ancilla state, $\rho_{ab}$, such that the ancilla ($b$) is subject to a local unitary evolution. In this picture, the ancilla acts as a measuring device for any operation performed on the system ($a$) (for an illustration, see Fig. \ref{fig:0}).
The system + ancilla Hamiltonian is given by $\mathcal{H}$ = $\mathcal{I}_a \otimes \mathcal{H}_b$, where $\mathcal{H}_b$ is the local Hamiltonian acting on $b$, and $\mathcal{I}_a$ is the identity operator acting on $a$. For any bipartite state, $\rho_{ab}$ and local Hamiltonian $\mathcal{H}_b$, the optimal available precision for the estimation of a parameter $\phi$ encoded in the local unitary $U_b = \exp(-i \phi \mathcal{H}_b)$, is governed by the quantum Fisher information \cite{QFI}, as derived using the Cram{\'e}r-Rao bound \cite{cram}, which is defined as follows.

For a quantum state, written in its spectral decomposition as  $\rho_{ab}$ = $\sum_m e_m |\phi_m\rangle\langle \phi_m|$, where $e_m \ge$ 0 and $\sum_m e_m$ = 1, the quantum Fisher information associated with the local evolution generated by $\mathcal{I}_a \otimes \mathcal{H}_b$ can be written as \cite{QFI},
\begin{equation}
\cal{F}(\rho_{ab},\mathcal{H}_b) = 4 \sum_{\substack{m,n: \\ e_m + e_n > 0}} \frac{(e_m - e_n)^2}{e_m + e_n} |\langle \phi_m| \mathcal{I}_a \otimes \mathcal{H}_b|\phi_n \rangle|^2.
\label{f}
\end{equation}
The above expression can be equivalently rewritten as follows,
\begin{equation}
\cal{F}(\rho_{ab},\cal{H}_b) = 4\textrm{tr}(\rho \cal{H}_b^2) - \sum_{\substack{m,n: \\ e_m + e_n > 0}} \frac{8e_m e_n}{e_m + e_n} |\langle \phi_m| \mathcal{I}_a \otimes \mathcal{H}_b|\phi_n \rangle|^2.
\label{qfi}
\end{equation}

If the generator  of the local evolution on the ancilla is not known {\it a priori}, as in the black-box paradigm for quantum metrology \cite{ades,GIP}, then the guaranteed precision enabled by the state $\rho_{ab}$ is given by the QIP (${\cal Q}$), defined as the minimum quantum Fisher information over all local Hamiltonians ${\cal H}_b$ of a fixed spectral class 
(a canonical choice is to consider the minimization to run over all ${\cal H}_b$ with nondegenerate, equispaced eigenvalues) \cite{comment0}, namely
\begin{equation}
\cal{Q}(\rho_{ab}) = \frac{1}{4} \inf_{{\cal H}_b} \cal{F}(\rho_{ab},\cal{H}_b)\label{qip}\,,
\end{equation}
where the $\frac14$ factor is a convenient normalization \cite{ades}.

In the relevant case where the system $a$ has arbitrary dimension, while the ancilla $b$ is a qubit, the choice of local Hamiltonians is reduced to  $\mathcal{H}_b$ = $\vec{r} \cdot \vec{\sigma}$, where $|\vec{r}|$ = 1 and $\vec{\sigma}$ = \{$\sigma^x,\sigma^y,\sigma^z$\} is the vector of the Pauli matrices \cite{ades, mb}. For such local Hamiltonians, the minimization in Eq.~(\ref{qip}) can be performed analytically, so that the QIP is computable in closed form and given by the expression \cite{ades,mb},
\begin{equation}
\cal{Q}(\rho_{ab})= 1 - \lambda_w^{max},
\label{lqf}
\end{equation}
where $\lambda_w^{max}$ is the highest eigenvalue of the real symmetric matrix $W$ with elements \cite{ades,mb},
\begin{equation}
W_{ij} = \sum_{\substack{m,n: \\ e_m + e_n > 0}} \frac{2e_m e_n}{e_m + e_n} \langle \phi_m| \cal{I}_a \otimes \sigma_b^i|\phi_n \rangle\langle \phi_n| \cal{I}_a \otimes \sigma^j_b|\phi_m\rangle.
\label{w}
\end{equation}

In general, the QIP is a {\it bona fide} measure of bipartite quantum correlations beyond entanglement, of the so-called discord type (see \cite{kav} for a review), in the quantum state $\rho_{ab}$. Namely,  $\cal{Q}(\rho_{ab})$ is known to vanish for states with zero discord from the perspective of subsystem $b$ (known as quantum-classical states), is invariant under local unitary operations, and reduces to an entanglement monotone for pure quantum states. Most importantly for the aims of the present paper, the QIP is a monotonically decreasing function under the action of arbitrary local CPTP maps on the system $a$
\cite{ades,mb}.
Furthermore, the QIP can be interpreted  as the minimal global speed of evolution for the state $\rho_{ab}$ under all local unitary transformations on the ancilla $b$, as a consequence of the connection between the quantum Fisher information and the Bures metric \cite{bures,buresdisc,mb}.

The evaluation of the QIP remains computationally tractable for higher-dimensional ($d_a \times d_b$) systems, although a closed analytical form may not be available. The problem can be recast in the form of a minimization of the Hamiltonian with respect to a finite number of variables spanning a compact space. This follows by noting that the unitary evolution, corresponding to $\cal{H}_b$ acting on the ancilla $b$, can be chosen within the special unitary group, without any loss of generality. Furthermore, the QIP can also be reliably computed for two-mode Gaussian states in the continuous variable regime \cite{GIP,Gmb}.

\subsection{Characterizing non-Markovianity}
\noindent

Let us consider an open quantum system undergoing an evolution given by the time-local master equation,
\begin{equation}
\frac{d}{dt} \rho(t) = \mathcal{L}(t)\rho(t),
\label{li}
\end{equation}
where $\mathcal{L}(t)$ is the Liouvillian superoperator \cite{lin1,lin2}, given by
\begin{eqnarray}
\mathcal{L}(t)\rho(t) &=& -i[\mathcal{H}(t),\rho(t)] - \sum_i \gamma_i(t)\bigg[A_i(t)\rho(t)A_i^\dag(t)- \nonumber\\
&&\frac{1}{2}\left\{A_i^\dag(t)A_i(t),\rho(t)\right\}\bigg].
\end{eqnarray}
Here  $A_i(t)$ are the Lindblad operators, and $\gamma_i(t)$ is the time-dependent relaxation rate.

The quantum evolution is Markovian when  $\gamma_i(t) \ge 0$ for each instant of time, $\forall\, t \ge 0$. The dynamical quantum process can be then defined in terms of time-ordered CPTP maps, such that
$
\Omega(t_2,t_1) = T \exp\left[\int_{t_1}^{t_2} dt' \mathcal{L}(t')\right],
$
where $T$ is the time-ordering operator. The map $\Omega(t,0)$ represents the evolution of the system from an initial state ($t=0$) to a state at time $t$.
Importantly, such CPTP map satisfies the divisibility criteria, in the sense that it can be written as a composition of other time-ordered CPTP maps, such that $\Omega(t + dt,0)$ = $\Omega(t + dt, t)\Omega(t,0)$.

Conversely, for instances where $\gamma_i(t) < 0$, the corresponding dynamical map $\Omega(t + dt, t)$ may not be CPTP and the divisibility property of the overall CPTP dynamics is violated. The non-divisibility of the dynamical maps given by a time-local master equation, of the form in Eq. (\ref{li}), is the essential marker of non-Markovian dynamics \cite{rep,arxiv,jpb} (cf. \cite{anti}).
As most quantum information quantities are monotonic under local CPTP maps, any observation of a nonmonotonic behavior in some reference quantity can be exploited to capture the breakdown of Markovianity.

In this work, we consider the QIP ($\mathcal{Q}(\rho_{ab})$)  as our reference information-theoretic figure of merit, and witness non-Markovianity in terms of the nonmonotonicity of the QIP under local evolutions of the system $a$, since for Markovian dynamics it must hold that  $\mathcal{Q}(\Omega_a\rho_{ab}) \le \mathcal{Q}(\rho_{ab})$, for all local CPTP maps $\Omega_a$ acting on the system $a$.

More precisely, to characterize non-Markovianity, let us begin with a basic description of the open system under consideration.
Let $a$ be the system interacting locally with an external environment and $b$ be the ancilla of the overall bipartite state $\rho_{ab}$.
We calculate the QIP in the bipartite system using the measure $\mathcal{Q}(\rho_{ab})$ defined in Eq.~(\ref{qip}), where the ancilla $b$ acts as the measuring apparatus on which the local unitaries are applied (see Fig. \ref{fig:0}). Now we consider the dynamics of the system $a$ as described by the time-local master equation of Eq.~(\ref{li}). For a Markovian dynamics, the evolution is given by a divisible CPTP map $\Omega_a(t)$. Using the Jamio{\l}kowski-Choi isomorphism, the composite dynamics of the overall system, $\rho_{ab}(t)$, is given by
\begin{equation}
\rho_{ab}(t) = \textbf{(}\Omega_a(t) \otimes \mathbb{I}_b\textbf{)}\rho_{ab}(0).
\label{dyn}
\end{equation}
Since the QIP is monotonically nonincreasing under local CPTP maps  acting on the system $a$, the function $\mathcal{Q}(\rho_{ab}(t))$ is monotonically nonincreasing with increasing time. Hence,
\begin{equation}
\frac{d}{dt} \mathcal{Q}(\rho_{ab}(t))\le 0
\end{equation}
holds for all $t \geq 0$ for a Markovian process. However, this may not be true for a non-Markovian process where the divisibility of the local CPTP map is violated. Therefore, $\frac{d}{dt} \mathcal{Q}(\rho_{ab}(t)) > 0$ is a straightforward non-Markovianity witness. If we define $\cal{D}(t) = \frac{d}{dt} \mathcal{Q}(\rho_{ab}(t))$, then the non-Markovianity of the dynamical map can be quantified using the expression,
\begin{equation}
\mathcal{N}_{\cal{Q}}(\Omega) = \max_{\rho_{ab}(0)} \int_{\cal{D}(t) > 0} \cal{D}(t) dt,
\label{N}
\end{equation}
where the maximization is performed over all sets of possible initial system + ancilla states, $\rho_{ab}(0)$. The integration is extended over all time intervals for which $\cal{D}(t) > 0$. Numerically, the final integration can be reduced to a summation of discrete sets of small interval integrals,
\begin{equation}
\mathcal{N}_{\cal{Q}}(\Omega) = \max_{\rho_{ab}(0)} \sum_{k: \cal{D}_k(t) > 0} \int_{t^k_i}^{t^k_f} \cal{D}_k(t) dt.
\label{NN}
\end{equation}
The maximization over all possible initial states involved in quantifying non-Markovianity is certainly demanding. However, starting with any chosen set of initial states, one can always obtain lower bounds to the non-Markovianity measure, thus achieving a qualitative assessment of the non-Markovian character of the dynamics.

\section{Qualitative analysis of non-Markovianity}
\label{non}
In order to qualitatively analyze the proposed non-Markovianity measure, $\cal{N}_\cal{Q}$, and compare it with some of the other important measures in the literature, we now consider the paradigmatic model of a single-qubit dephasing channel. The Hamiltonian describing a single qubit interacting with a thermal reservoir is given by \cite{open1},
\begin{equation}
\cal{H}= \omega_0\sigma^z + \sum_i \omega_i \hat{a}_i^\dag \hat{a}_i + \sum_i (g_i \sigma^z \hat{a}_i + g^*_i \sigma^z \hat{a}^\dag_i),
\label{h1}
\end{equation}
where $\omega_0$ is the qubit resonant transition frequency, $\hat{a}_i$ ($\hat{a}_i^\dag$) and $\omega_i$ are the annihilation (creation) operators and frequency of the $i$th reservoir mode, and $g_i$ is the reservoir-qubit coupling constant for each mode. The qubit dynamics resulting from the Hamiltonian in Eq. (\ref{h1}) is given by the differential equation,
\begin{equation}
\dot{\rho}(t) = \gamma(t)(\sigma^z \rho(t) \sigma^z - \rho(t)),
\end{equation}
where $\gamma(t)$ is the time-dependent decoherence rate, which can be determined from the spectral density ($J(\omega)$) of the coupling constants \cite{comment,J}, and $\sigma^z$ is the third Pauli operator. The dynamical map for the dephasing channel ($\Delta(t)$), on a single-qubit system ($\rho_a$) is given by
\begin{equation}
\rho_a(t)= \Delta(t)\rho_a(0)=\\
\left(\begin{array}{cc}
{\rho_a}_{00}(0) & {\rho_a}_{10}(0) \Gamma(t)       \\ \\
{\rho_a}_{01}(0)\Gamma(t) & {\rho_a}_{11}(0)
\end{array}\right),
\label{de}
\end{equation}
where $\Gamma(t) = \exp \left[-2\int_0^t \gamma(t')dt'\right]$, and ${\rho_a}_{ij}(0)$ are the elements of the initial system state, $\rho_a(0)$. For a zero-temperature reservoir with spectral density $J(\omega)$, the decoherence rate is given by the relation \cite{open1,range},
\begin{equation}
\gamma(t) = \int J(\omega)\frac{\sin(\omega t)}{\omega} d\omega.
\label{j}
\end{equation}

To analyze the non-Markovianity, we need to calculate the measure $\mathcal{N}_{\cal{Q}}$, given by Eq. (\ref{N}), for the composite system + ancilla state, $\rho_{ab}$ (where the system undergoes dephasing while the ancilla is not subject to decoherence) optimized over all possible initial states $\rho_{ab}(0)$. Such an optimization process is complicated and can be solved only for specific instances. Alternately, a lower bound on $\mathcal{N}_{\cal{Q}}$ can be obtained  by considering
the particular situation where the initial composite system + ancilla state, $\rho_{ab}(0)$, is maximally entangled, say  $\rho_{ab}(0) = |\Phi\rangle\langle\Phi|$, where $|\Phi\rangle$ is a Bell state, $|\Phi\rangle = (|00\rangle + |11\rangle)/\sqrt2$.

The composite dynamical map is given by \cite{LUO},
\begin{equation}
\rho_{ab}(t)= (\Delta_a(t) \otimes \mathbb{I}_b)|\Phi\rangle\langle\Phi|=\\
\frac12 \left(\begin{array}{cccc}
1 & 0 & 0 & \Gamma(t) \\
0 & 0 & 0 & 0 \\
0 & 0 & 0 & 0 \\
\Gamma(t) & 0 & 0 & 1
\end{array}\right),
\label{dej}
\end{equation}
and the non-optimized measure of non-Markovianity is given by the relation,
\begin{equation}
\mathcal{N}^{~0}_{\cal{Q}}(\Delta_a(t)) =  \int_{\substack{\cal{D}(t) > 0\\ \rho_{ab}(0)=|\Psi\rangle\langle\Psi|}} \cal{D}(t) dt.
\label{No}
\end{equation}

We need to calculate the QIP in the evolved state $\rho_{ab}(t)$ using the measure $\cal{Q}$, given by Eqs.~(\ref{qip})--(\ref{w}). For the state $\rho_{ab}(t)$, given by Eq.~(\ref{dej}), the maximum eigenvalue of the $W$ matrix, $\lambda_w^{max}$, is equal to $1 - \Gamma(t)^2$. Hence,
\begin{eqnarray}
&\cal{Q}(\rho_{ab}(t)) = \sqrt{1 - \lambda_w^{max}} = \Gamma(t)& \\
&\cal{D}(t) = \frac{d}{dt} \mathcal{Q}(\rho_{ab}(t)) = - 2 ~\Gamma(t) \gamma(t).&
\end{eqnarray}
Since $\Gamma(t) > 0$ for all $t$, it follows that  $\cal{D}(t) > 0$ when $\gamma(t) < 0$, which is consistent with other well-established definitions of non-Markovianity \cite{RHP,BLP,LUO}. The measure of non-Markovianity based on QIP for the initially maximally entangled, composite system + ancilla state, $\rho_{ab}$, under single-qubit dephasing on the system $a$, is given by
\begin{equation}
\mathcal{N}^{~0}_{\cal{Q}}(\Delta_a(t)) =  -2 \int_{\gamma(t)< 0} \Gamma(t) \gamma(t) dt.
\end{equation}
Interestingly, the quantification of non-Markovianity in terms of QIP ($\cal{Q}$), for the paradigmatic single-qubit dephasing model and maximally entangled initial states, is numerically equivalent to the previously introduced measure in terms of the distinguishability of a pair of evolving states using the trace distance \cite{BLP}. The distinguishability witness for non-Markovianity is closely associated with the backflow of information from the environment to the system, which results in the increase of quantum correlations in the dephased bipartite state $\rho_{ab}$ as detectable through the QIP. Hence, the non-Markovianity defined in terms of the local quantum Fisher information, whose minimization defines the QIP, exactly captures the intrinsic backflow of information in the system-environment interaction.

To further compare the measure $\mathcal{N}^{~0}_{\cal{Q}}$ against other measures of non-Markovianity, we consider specifically the above single-qubit dephasing model with an Ohmic reservoir spectral density,
\begin{equation}
J(\omega) = \alpha\omega_c \left(\frac{\omega}{\omega_c}\right)^S \exp\left(-\frac{\omega}{\omega_c}\right),
\end{equation}
where $S$ is the Ohmicity parameter, $\alpha$ is the dimensionless coupling constant, and $\omega_c$ is the cutoff spectral frequency. For a zero reservoir temperature, relation (\ref{j}) can be written as,
\begin{equation}
\gamma(t) = \alpha \omega_c \frac{\cos\left[S \tan^{-1}(\omega_c t)\right] \Gamma_0(S)}{(1+\omega_c^2 t^2)^{S/2}},
\label{g}
\end{equation}
where $\Gamma_0( \cdot )$ is the Euler Gamma function. It is known that the dephasing dynamical map corresponding to the spectral density $J(\omega)$ is divisible in the parameter range $0 < S \le 2$ \cite{range}. Hence, the non-Markovian regime corresponds to the super-Ohmic parameter range $S >$ 2, which can be experimentally obtained in ultracold systems utilizing control over atomic noise \cite{capacity, J}. To calculate $\mathcal{N}^{~0}_{\cal{Q}}$, we need to integrate over the relevant range of $t$, where $\gamma(t) < 0$ and $\cal{D}(t) > 0$.

\begin{figure}
\centering
\includegraphics[width=0.48\textwidth, angle=0]{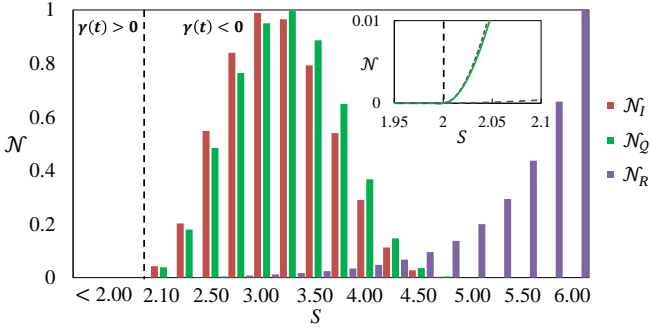}
\caption{\label{fig:1}(Color online). In this figure, we compare the  measure $\cal{N}_\cal{Q}$ (= $\mathcal{N}^{~0}_{\cal{Q}}$), with other quantifiers of non-Markovianity, for a single-qubit dephasing channel with an Ohmic spectral density, $J(\omega)=\alpha\omega_c \left(\omega/\omega_c\right)^S \exp\left(\omega/\omega_c\right)$.
The blue dashed region, corresponding to $\gamma(t) >$ 0, is the Markovian regime. This corresponds to $\mathcal{N}$ = 0, for all known measures of non-Markovianity. The region $\gamma(t) <$ 0 denotes  the non-Markovian regime and we plot the behavior of the measures based on QIP ($\mathcal{N}_\cal{Q}$, green bar), quantum mutual information ($\mathcal{N}_I$,  red bar), and divisibility criteria ($\mathcal{N}_R$, purple bar). We observe that the non-Markovian regime lies in the super-Ohmic region, $S > 2$.
The behavior of the measures, $\mathcal{N}_\cal{Q}$ and $\mathcal{N}_I$, is similar: Both indicators  initially increase, followed by a decrease, with increasing $S$, with vanishing values for $S >$ 5. However, in contrast the measure $\mathcal{N}_R$ monotonically increases with $S$ in the non-Markovian region. This implicitly shows that the measures based on QIP and mutual information are independent of the divisibility criteria. The inset figure shows the measures of non-Markovianity around the critical transition parameter value, $S$ = 2. We note that, for this model, the non-Markovianity measure based on distinguishability is identical to the one derived in this work. }
\end{figure}

Figure \ref{fig:1} compares different measures of non-Markovianity based on QIP, quantum mutual information \cite{LUO}, distinguishability \cite{BLP}, and divisibility criteria \cite{RHP}. For a set of maximally entangled initial states, we observe that the measures based on the first three quantities behave in a similar manner as opposed to the non-Markovian measure based on divisibility. The measures based on QIP and distinguishability are equivalent for this model.

\section{Flow of quantum interferometric power in non-Markovian dynamics}
\label{flow}

\noindent We now consider the flow of the QIP ($\cal{Q}$) in the dissipative dynamics governed by the single-qubit amplitude damping channel. The qubit dynamics can be modelled using the Hamiltonian given by \cite{open1},
\begin{equation}
\cal{H}= \omega_0\sigma^z + \sum_i \omega_i \hat{a}_i^\dag \hat{a}_i + \sum_i (g_i \sigma^+ \hat{a}_i + g^*_i \sigma^- \hat{a}^\dag_i),
\label{h2}
\end{equation}
where, $\sigma^+$ ($\sigma^-$) is the raising (lowering) Pauli operator. The resulting dynamics is given by the differential equation,
\begin{equation}
\dot{\rho} = -i\frac{s(t)}{2}[\sigma^+\sigma^-,\rho] + \gamma(t)\left(\sigma^-\rho\sigma^+ - \frac{1}{2}\left\{\sigma^+\sigma^-,\rho \right\}\right),
\end{equation}
where $\rho$ is time dependent. The functions $s(t)$ and $\gamma(t)$ are defined in terms of the integro-differential equation for the time-dependent function, $\cal{J}_t$, given by
\begin{equation}
\dot{\cal{J}}_t = - \int_0^t d\tau~ f(t-\tau)\cal{J}_t,
\end{equation}
where, $s(t)$ = --2Im$\frac{\dot{\cal{J}}_t}{\cal{J}_t}$ and $\gamma(t)$ = --2Re$\frac{\dot{\cal{J}}_t}{\cal{J}_t}$. The function $\cal{J}_t$ is characteristic of the nature of the environment used to model the local noise in the dynamics.  It is defined in terms of the correlation function $f(t-\tau)$, that is derived from the Fourier transform of the spectral density of the environment, $J(\omega)$:
\begin{equation}
f(t-\tau)=\int  J(\omega) \ e^{i(\omega_0-\omega)(t-\tau)} d \omega.
\end{equation}
The dynamics of the system qubit  ($\rho_a$), is given by the dynamical map $\rho_a(t)= \Omega(t)\rho_a(0)$, such that \cite{open1,openex1}
\begin{equation}
\rho_a(t)=\\
\left(\begin{array}{cc}
{\rho_a}_{00}(0)+{\rho_a}_{11}(0)(1-\cal{J}_t^2) &~ {\rho_a}_{01}(0) \cal{J}_t       \\ \\
{\rho_a}_{10}(0)\cal{J}_t^{*} &~ {\rho_a}_{11}(0)\cal{J}_t^2
\end{array}\right).
\label{da}
\end{equation}
The single-qubit amplitude damping dynamics can be extended to the system + ancilla bipartite system. In the computational basis, the evolved density matrix of the composite two-qubit system $\rho_{ab}(t)$ is given as follows (omitting the subscript $ab$ for simplicity). The diagonal matrix elements of $\rho_{ab}$ are given by \cite{qmb,dis1}
\begin{eqnarray}
&\rho_{11,11}(t)=\rho_{11,11}(0)\cal{J}_t^2; \rho_{10,10}(t)= \rho_{10,10}(0)\cal{J}_t^2 ;&\nonumber\\
&\rho_{01,01}(t)=\rho_{01,01}(0) + \rho_{11,11}(0) (1-\cal{J}_t^2);&\nonumber\\
&\rho_{00,00}(t)=1 - (\rho_{01,01}(t)+\rho_{10,10}(t)+\rho_{11,11}(t)).&\nonumber
\end{eqnarray}
The nondiagonal elements are given by
\begin{eqnarray}
&\rho_{11,10}(t)=\rho_{11,10}(0)\cal{J}_t^2; \rho_{11,01}(t)= \rho_{11,01}(0)\cal{J}_t;&\nonumber\\
&\rho_{11,00}(t)= \rho_{11,00}(0)\cal{J}_t; \rho_{10,01}(t)= \rho_{10,01}(0)\cal{J}_t;&\nonumber\\
&\rho_{01,00}(t)=\rho_{01,00}(0) + \rho_{11,10}(0)(1-\cal{J}_t^2);&\nonumber\\
&\rho_{10,00}(t)= \rho_{10,00}(0)\cal{J}_t,&\nonumber
\end{eqnarray}
where $\rho_{i,j}=\rho_{j,i}^*$ ($i,j={00,01,10,11}$).

Let us now consider a reservoir spectral density with a Lorentzian distribution \cite{open1,open2},
\begin{equation}
J(\omega)= \frac{\gamma_0\lambda^2}{2\pi[(\omega-\omega_c)^2+\lambda^2]},
\end{equation}
where $\omega_c$ is the central frequency of the distribution and $\gamma_0$ is the system-reservoir coupling constant. The spectral width of the distribution, $\lambda$, is the inverse of the reservoir correlation time $(\tau_r = \frac{1}{\lambda})$. The system-reservoir coupling $\gamma_0$ is related to the Markovian decay of the system, and is thus the inverse of the system relaxation time $(\tau_s = \frac{1}{\gamma_0})$.
%
The Markovian nature of the dynamics is related to the strength of the system-reservoir coupling and the interplay of the system relaxation and reservoir correlation times. For weak coupling, the relaxation time of the system is greater than the reservoir correlation time, $\tau_s > 2\tau_r$ ($\gamma_0 < \frac{\lambda}{2}$), and the dynamics is essentially Markovian. For, $\tau_s < 2\tau_r$ ($\gamma_0 > \frac{\lambda}{2}$) or in the strong coupling regime, the dynamics is non-Markovian. Hence, the non-Markovian character of the considered dynamical map is ingrained in the behavior of the function $\cal{J}_t$, which for a Lorentzian spectral distribution is of the form \cite{open1,openex1},
\begin{equation}
\cal{J}_t=e^{\frac{-(\lambda-i\delta)t}{2}}\left[\mbox{cosh}\left(\frac{\eta t}{2}\right) + \frac{(\lambda-i\delta)}{\eta}\mbox{sinh}\left(\frac{\eta t}{2}\right)  \right],
\label{jt}
\end{equation}
where $\eta=\sqrt{(\lambda-i\delta)^2-2\gamma_0 \lambda}$, and $\delta$ (= $(\omega_0-\omega_c)$) is the system-reservoir frequency detuning. For the dynamics to be Markovian, in the weak coupling regime, the function $\cal{J}_t$ needs to have a monotonic decrease with time. For non-Markovian dynamics, the monotonicity of $\cal{J}_t$ does not hold, consistent with the breakdown of the divisibility of the dynamical CPTP map.


Now, let us consider our initial bipartite state to be a Bell-diagonal Werner state of the form,
\begin{equation}
\rho_{ab}(0) = \frac{1}{4}\left(\mathbb{I}_4+\sum_{i=x,y,z} r_i \sigma^i \otimes \sigma^i\right),
\label{wer}
\end{equation}
where the Werner parameter is given by, $r_x = -r_y = r_z = r \in [0,1]$. The initial state  $\rho_{ab}(0)$ is maximally mixed for $r=0$, and a maximally entangled Bell state for $r=1$.
The resulting time-evolved density matrix, under a single-qubit amplitude damping, can be written as
\begin{equation}
\rho_{ab}(t)=\\
 \frac{1}{4}\left(\begin{array}{cccc}
\bar{r}_+\cal{J}_t^2 & 0 & 0 &  2 r \cal{J}_t       \\
0 & \bar{r}_-\cal{J}_t^2 & 0 &  0       \\
0 & 0 & \bar{r}_{-} + \bar{r}_+\bar{\cal{J}_t^2} &  0       \\
2 r \cal{J}^*_t & 0 & 0 & \bar{r}_+ +\bar{r}_-\bar{\cal{J}_t^2}
\end{array}\right),
\label{dis}
\end{equation}
where, $\bar{r}_\pm$ = 1 $\pm~ r$, and $\bar{\cal{J}_t^2}$ = 1 $-\cal{J}_t^2$. Using Eqs. (\ref{lqf}), (\ref{w}), and (\ref{jt}), the dynamical flow of QIP ($\mathcal{Q}$)
can be numerically evaluated for the density matrix given by Eq. (\ref{dis}). The non-Markovian character of the local dynamics of the bipartite qubit system $ab$ can be analyzed by observing the evolution of $\mathcal{Q}(\rho_{ab}(t))$, with increasing time $t$ of evolution.

We begin by analyzing the case of a maximally entangled initial state $\rho_{ab}(0)$, obtained by setting $r=1$ in Eq.~(\ref{wer}). If we consider $\cal{J}_t$ to be a complex number of the form, $\cal{J}_t=\alpha_t + i~ \beta_t$, where $\alpha_t$, $\beta_t \in \mathbb{R}~ \forall~ \alpha_t$, $\beta_t \in \left[0,1\right]$,  the maximum eigenvalue $\lambda_w^{max}$ of the matrix $W$, Eq.~(\ref{w}), is equal to $(1-\alpha_t^2-\beta_t^2)$. Hence, the QIP $\mathcal{Q}(\rho_{ab})$ is given by,
\begin{eqnarray}
\mathcal{Q}(\rho_{ab}(t)) = \sqrt{1-\lambda_w^{max}}
=\sqrt{\alpha_t^2+\beta_t^2} = |\cal{J}_t|.
\label{equal}
\end{eqnarray}
Therefore, for a maximally entangled initial state undergoing a single-qubit amplitude damping, the nonmonotonic flow of the QIP in the non-Markovian regime is exactly governed by the nonmonotonicity of the function $\cal{J}_t$ (cf. \cite{dis1}). Interestingly,  $|\cal{J}_t|$ also measures the maximal trace distance between a pair of system states and hence quantifies the non-Markovianity in terms of distinguishability for single-qubit amplitude damping channels \cite{feng}. We find therefore that the flow of the QIP  $\mathcal{Q}(\rho_{ab}(t))$ is again closely related to the backflow of quantum information in the non-Markovian regime.

\begin{figure}
\centering
\includegraphics[width=0.42\textwidth, angle=0]{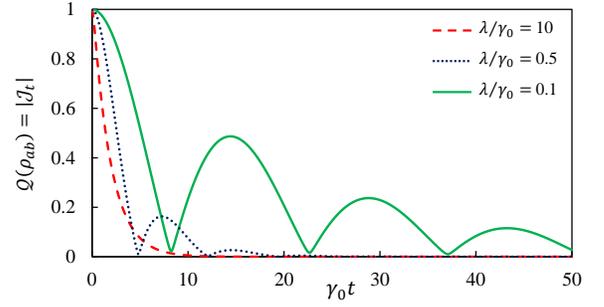}
\caption{\label{fig:2}(Color online) The flow of QIP ($\cal{Q}$) in a two-qubit system with maximal entanglement at $t$=0 under a single-qubit amplitude damping channel. The initial entanglement is set by setting $r$ = 1 in the Werner state given by Eq. (\ref{wer}). $\mathcal{Q}(\rho_{ab})$ is equal to $|\cal{J}_t|$, as shown by Eq. (\ref{equal}).
We observe the flow of $\cal{Q}$ in both Markovian and non-Markovian regimes of the dynamical evolution. The flow under the Markovian regime (red dashed line) corresponds to the ratio of the reservoir correlation to system relaxation, $\lambda/\gamma_0$ = 10. Under the non-Markovian regime, the flow is shown for $\lambda/\gamma_0$ = 0.5 (blue dotted line) and $\lambda/\gamma_0$ = 0.1 (green solid line). The system-reservoir frequency detuning is set at, $\delta$ = 0.01 $\gamma_0$. The backflow of quantum correlation, in terms of QIP, is observed by the nonmonotonic increase of $\cal{Q}$ during the evolution. }
\end{figure}

Figure \ref{fig:2} shows the flow of the QIP measure for an initial maximally entangled state, given by Eq.~(\ref{wer}), with $r$ =1. The Markovian and non-Markovian regimes of the dynamics can be studied in terms of the function $\cal{J}_t$, for a Lorentzian reservoir spectral distribution, as mentioned in Eq. (\ref{jt}). The Markovian regime corresponds to $\gamma_0/\lambda <$ 0.5, as shown in the figure for $\lambda/\gamma_0$ = 10. The non-Markovian regime, corresponding to strong system-reservoir coupling $\gamma_0/\lambda >$ 0.5, is shown for $\lambda/\gamma_0$ = 0.1 and 0.5. The system-reservoir detuning is $\delta$ = 0.01 $\gamma_0$. The figure shows that the non-Markovian flow of QIP is nonmonotonic, with increase in $\cal{Q}$ during certain evolution times. The non-Markovianity can be numerically evaluated using the expression for $\cal{N}_\cal{Q}$ in Eqs. (\ref{N}) and (\ref{NN}).

\begin{figure}[t]
\centering
\includegraphics[width=0.4\textwidth, angle=0]{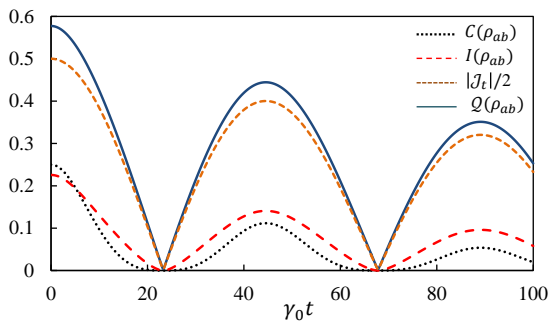}
\caption{\label{fig:3}(Color online). The flow of QIP ($\cal{Q}$) for an initially mixed two-qubit system under a single qubit amplitude damping channel. The initial mixed state is obtained for $r$ = 0.45, in the Werner state given by Eq. (\ref{wer}).The parameters pertaining to the Lorentzian reservoir spectral distribution are set at $\lambda = 0.01 \gamma_0$, and $\delta = 0.001 \gamma_0$. The figure shows the non-Markovian evolution of  $\cal{Q}(\rho_{ab})$ (blue solid line), concurrence, $C(\rho_{ab})$ (black dotted line), quantum mutual information, $I(\rho_{ab})$ (red broad-dashed line), and the scaled function $|\cal{J}_t|/2$ (brown dashed line).
}
\end{figure}

We have seen that the flow of QIP for a maximally entangled initial state is determined by the nature of spectral distribution and is equal to the integro-differential function $|\cal{J}_t|$. However, the situation is not so straightforward if the initial system + ancilla state is mixed. This can be obtained easily from the Werner state in Eq. (\ref{wer}), by setting the Werner parameter $r <$ 1. In such instances, the flow of QIP is still governed by the  (non-)monotonic behavior of $|\cal{J}_t|$. We observe that the QIP ($\cal{Q}$) has a dynamical behavior that quite closely replicates the flow of the function $|\cal{J}_t|$, as compared to other measures such as the quantum mutual information \cite{mut} or the entanglement (quantified by the concurrence) \cite{concur}. Figure \ref{fig:3}, shows the dynamics of quantum information measures in the case of an initially mixed Werner state $\rho_{ab}$, defined for $r$ = 0.45, in the non-Markovian regime of the single-qubit amplitude damping channel. The reservoir relaxation is set at $\lambda = 0.01 \gamma_0$, and the detuning is $\delta = 0.001 \gamma_0$. The figure shows specifically the flow of QIP, concurrence, and the mutual information in comparison to the behavior of the scaled function $|\cal{J}_t|/2$. For the considered model, the behavior of $\cal{Q}(\rho_{ab})$ closely follows the nonmonotonic and discontinuous evolution of the function $\cal{J}_t$. This is in contrast to entanglement and mutual information, which both evolve smoothly with time. Furthermore, entanglement decays quickly and vanishes for finite ranges of time (so-called entanglement sudden death \cite{esd}) and hence cannot qualitatively capture the backflow of quantum information in selected intervals of time.
%

\section{\label{sec:Conc} Discussion and conclusion}

Non-Markovianity is an ubiquitous feature of quantum dynamical maps, and is nowadays recognized as a resource for certain applications of quantum technology, such as metrology, cryptography, and communication \cite{qkd,qprot,qmet,capacity}.
The role of non-Markovianity in enhancing the robustness of quantum correlations in systems exposed to noisy environments has been studied by means of various quantitative approaches \cite{RHP,LUO,arxiv}. In this work, we have adopted the quantum interferometric power (QIP) \cite{ades,mb} as our reference figure of merit to assess non-Markovianity of dynamical maps applied to a system coupled to an ancilla, which plays the role of a measuring apparatus for the operations occurring on the system. The QIP has been very recently acknowledged as a physically insightful, operationally motivated, and computable measure of quantum correlations of the most general kind, including and beyond entanglement \cite{ades,golden}. The QIP corresponds to the guaranteed metrological precision that a system + ancilla probe state enables the estimation of a phase shift on the ancilla part, in the black-box quantum metrology paradigm \cite{ades,mb}. When the system is subject to a non-Markovian evolution, the QIP between system and ancilla (measured from the perspective of the ancilla) can undergo a nonmonotonic evolution, with revivals in time. We have shown that such nonmonotonic behavior is closely related to, and can precisely capture, the backflow of information from environment to system which is a clear marker of non-Markovianity \cite{BLP}. In operative terms, such a dynamical rise of quantum correlations translates into an increase of the guaranteed precision of phase  estimation on the ancilla, thanks to the non-Markovian noise affecting the system.
While here we considered paradigmatic dynamical maps applied to single-qubit systems only, it has been shown in \cite{qmet} that non-Markovian noise affecting a register of $n$ qubits can lead to an enhancement in the metrological scaling which is intermediate between the shot noise and the Heisenberg limit. It will be interesting to investigate how the measure of non-Markovianity proposed here in terms of QIP can be employed to investigate the metrological scaling in the black-box paradigm for quantum metrology with multiqubit probes.

In this work we have proposed to quantify non-Markovianity in terms of the nonmonotonicity of the QIP, similarly to previous proposals to quantify non-Markovianity in terms of the nonmonotonicity of entanglement or total correlation measures \cite{RHP,LUO}. By analyzing two simple models of single-qubit noisy dynamics, we have shown how our measure reliably captures the non-Markovian regime, and is quantitatively more sensitive than measures based on entanglement. Another advantage associated with the use of the QIP to characterize non-Markovianity, is that such a measure also has been extended to Gaussian states of continuous variable systems, resulting in a computable and reliable measure of quantum correlations in that relevant setting as well \cite{GIP,Gmb}. In a subsequent work, it will be worth analyzing non-Markovianity in Gaussian dynamical maps \cite{gaus,gaus2} in terms of the nonmonotonic flow of the Gaussian QIP \cite{GIP}. The QIP has therefore the potential to offer a unified picture of non-Markovianity extending from qubits to infinite-dimensional systems.

We hope that the present analysis can stimulate further research in order to pin down the relevance of non-Markovian dynamics in quantum information processing and in the description and simulation of complex quantum systems in the biological, physical, and social domains \cite{CYB}.

\acknowledgments{
We are grateful to  Aditi Sen(De) and Ujjwal Sen for fruitful discussions. H.S.D. thanks the University Grants Commission (India) for support during doctoral research. G.A. thanks the ERC StG GQCOP (Grant~No.~637352), the Foundational Questions Institute (Grant~No.~FQXi-RFP3-1317), and the Brazilian CAPES (Grant~No.~108/2012) for financial support. G.A. acknowledges the kind hospitality of the Harish-Chandra Research Institute, Allahabad, India, during a visit.}


\begin{thebibliography}{99}
\bibitem{comment0} The Hamiltonian $\cal{H}_b$ belongs to a set of local Hamiltonians, acting on the ancilla $b$, with a fixed nondegenerate eigenvalue spectrum. For every chosen spectrum, by minimization of the Fisher information over the corresponding set of Hamiltonians one obtains a QIP measure. Here, we can focus on the canonical choice of Hamiltonians $\cal{H}_b$ with nondegenerate and equispaced eigenvalues. In this way, the unitary shift acting on the ancilla $b$ reduces to a root-of-unity unitary operation.  See Ref. \cite{ades} for details.
\bibitem{comment} The coupling constants in a continuous distribution correspond to a reservoir spectral distribution $J(\omega)$, such that, $\sum_i |g_i|^2 \rightarrow \int J(\omega) \delta(\omega_i - \omega) d\omega$. See Refs. \cite{open1,capacity,J} for details.
\end{thebibliography}
\end{document}